\documentclass[aps,showpacs,amsmath,amssymb,prl,twocolumn]{revtex4}
\usepackage{pslatex}
\begin{document}

\title{The divergence of neighboring magnetic field lines and
fast-particle diffusion in strong magnetohydrodynamic turbulence,
with application to thermal conduction in galaxy clusters}
\author{Jason Maron and Benjamin D. G. Chandran}
\affiliation{Department of Physics and Astronomy, University of Iowa}
\email{jason-maron@uiowa.edu, benjamin-chandran@uiowa.edu}
\author{Eric Blackman}
\affiliation{Department of Physics and Astronomy, University of Rochester}
\email{blackman@pas.rochester.edu }

\begin{abstract}
Using direct numerical simulations, we calculate the rate
of divergence of neighboring magnetic field lines in different
types of strong magnetohydrodynamic turbulence. 
In the static-magnetic-field approximation,
our results imply that tangled magnetic fields in galaxy
clusters reduce the electron diffusion coefficient and
thermal conductivity by a factor of~$\sim 5-10$
relative to their values in a non-magnetized plasma.
\end{abstract}
\maketitle
                      
The diffusion of fast particles in turbulent magnetized plasmas is
important for fusion experiments, cosmic-ray propagation, and thermal
conduction in galaxy-cluster
plasmas~\cite{fab94,bin81,ros89,tri89,chu93,tao95,pis96,cha98,mal01,jok73,ski74,nar01,gru02}.
We consider particle diffusion in a static magnetic field, which is a
reasonable first approximation for particles moving much faster than
the ${\bf E} \times {\bf B} $ velocity of field lines.  The effects of
turbulent bulk motions and field evolution have recently been
considered by~\cite{gru02,cha03}.

We consider magnetic fluctuations that possess an inertial
range extending from an outer scale $l_B$ to a much smaller inner
scale $l_d$ with the magnetic energy dominated by scales $\sim
l_B$. Except where specified, the discussion focuses on the case in
which there is no mean magnetic field pervading the plasma.  If a
particle is tied to a single field line and travels a distance $l\gg
l_B$ along the static magnetic field, it takes $\sim l/l_B$ random-walk
steps of length $l_B$, resulting in a mean-square three-dimensional
displacement of
\begin{equation}
\langle (\triangle x)^2 \rangle = \alpha l_B l,
\label{eq:tr} 
\end{equation} 
where $\alpha$ is a constant of order unity. The values of $\alpha$
for the turbulence simulations used in this paper are listed in
table~\ref{tab:t1}. When there is a mean field ${\bf B} _0$ comparable to
the rms field, $\langle (\triangle x)^2 \rangle$ in equation~(\ref{eq:tr}) 
is interpreted as the mean-square displacement
perpendicular to ${\bf B}_0$. If the particle's motion along the
field is diffusive with diffusion coefficient $D_\parallel$, then
\begin{equation}
l \sim \sqrt{D_\parallel t},
\label{eq:l1} 
\end{equation} 
and~\cite{rec78,kro83}
\begin{equation}
\langle (\triangle x)^2 \rangle \propto t^{1/2},
\label{eq:t2} 
\end{equation} 
indicating subdiffusion [$D\equiv \lim_{t \rightarrow \infty} \langle(\triangle x)^2
\rangle/6t \rightarrow 0$].

In fact, a particle is not tied to a single field line. After
traveling a short distance along the field, field gradients and scattering cause a
particle to take a step comparable to its gyroradius $\rho$ across the
magnetic field, from its initial field line,~F1, to a new field line,~F2. 
If the particle were to follow~F2, it would separate from F1 because
neighboring field lines tend to diverge. We call $z_s$ the
distance the particle would have to follow F2 before separating from F1
by a distance $l_B$.  (A particle typically separates from its initial
field line by a distance $l_B$ after traveling a distance slightly
less than $z_s$ since it drifts across the field continuously, but we ignore
this effect in this paper.)

If a particle ``takes a random step'' of length~$mz_s$ along the
magnetic field, where $m$ is a constant of order a few, reverses
direction, and then takes another random step of length~$mz_s$ back
along the field, it doesn't return to its initial point. Part of the
second step retraces part of the first step, but the remainder is
uncorrelated from the first step.  This loss of correlation leads to a
Markovian random walk~\cite{cha98,cha03}.
When $mz_s \gg l_B$, a single random step
corresponds to a 3D displacement of
\begin{equation}
(\triangle x)^2 \sim \alpha mz_s l_B.
\label{eq:tr3} 
\end{equation} 
When $mz_s \gg \lambda$, where $\lambda$ is the Coulomb mean free path,
a single step takes a time
\begin{equation}
\triangle t  \sim \frac{ m^2 z_s^2}{D_\parallel}.
\label{eq:dt} 
\end{equation} 
When $mz_s$ is only moderately greater than $l_B$ or $\lambda$, equations~(\ref{eq:tr3}) 
and (\ref{eq:dt}) remain approximately valid. During successive
random steps, a particle will find itself in regions of differing
magnetic shear, and thus $z_s$ will vary. The diffusion coefficient
is given by $D= \langle (\triangle x)^2 \rangle/6\langle \triangle t \rangle$
where $\langle \dots \rangle$ is an average over a large number of steps~\cite{cha43}.
Ignoring factors of order unity,
\begin{equation}
D \sim D_\parallel \frac{l_B}{L_{\rm S}}
\label{eq:d2} 
\end{equation} 
as in~\cite{rec78,cha98,cha03}, with
\begin{equation}
L_{\rm S} = \frac{\langle z_s^2\rangle}{\langle z_s \rangle}.
\label{eq:deflrr} 
\end{equation} 
If there is a mean magnetic field comparable to the fluctuating field,
equation~(\ref{eq:d2}) is recovered provided $D$ is replaced by
$D_\perp$, the coefficient of diffusion perpendicular to the mean
field.

We treat field-line separation in a static field using magnetic-field
data obtained from four direct numerical simulations of incompressible
MHD turbulence. Key simulation parameters are given in
table~\ref{tab:t1}. The numerical method is described in~\cite{mar01}.
Each simulation uses Newtonian viscosity~$\nu$ and resistivity~$\eta$ and is run
until a statistical steady state is reached.  In simulations~A1
and~A2, the mean magnetic field is zero, and the initial magnetic
field is dominated by large-scale fluctuations containing 10\% of the
maximum possible magnetic helicity. Turbulent fluctuations are
sustained by non-helical random forcing of the velocity field,
and the magnetic Prandtl number~$P_m\equiv\nu/\eta$ equals~1.  These
simulations reach a statistical steady state with Kolmogorov-like
kinetic and magnetic power spectra.  In simulation~B, $P_m=75$ and the field is
amplified from an initially weak seed magnetic field by turbulent
velocities sustained by non-helical random forcing.
When the dynamo growth of the magnetic field saturates, the 
magnetic field is dominated by fluctuations on scales much smaller than
the dominant velocity length scale, as in the simulations of~\cite{marb}. In simulation~D,
$P_m =1$ and the field is
amplified from an initially weak seed magnetic field by turbulent
velocities sustained by maximally helical random forcing. When the
dynamo growth of the magnetic field saturates, the dominant magnetic
field length scale is comparable to the dominant velocity length scale,
as in the simulations of~\cite{mara}. ``Simulation'' A2$_{\rm rp}$ is
obtained by assigning each Fourier mode in simulation~A2 a
random phase without changing the mode's amplitude. In simulations~A1,
A2, A2$_{\rm rp}$, and~D we set~$\pi/l_B$ equal to the
maximum of $k E_b(k)$, where $E_b(k)$ is the power
spectrum of the magnetic field [the total magnetic energy is $\int
E_b(k) dk$]. In simulation~B we take $\pi/l_B$ to be the maximum of
$kE_v(k)$, where $E_v(k)$ is the  power spectrum of the
velocity field.  We set $\pi/l_d$ equal to the maximum of $k^3E_b(k)$.

\begin{table}[h]
\begin{tabular}{ccccccccc}
\hline
\hline 
&&&&&&
\vspace{-0.35cm} 
\\
Simulation & Grid points & $\displaystyle
\frac{\delta B}{|\langle{\bf B}\rangle|}$ & $H_m$ & $l_B/l_d$ & $\alpha$  & \hspace{0.1cm} $P_m$
\hspace{0.1cm}  \vspace{0.05cm} \\
\hline  
\vspace{-0.4cm} \\
A1 & $256^3$ & $\infty$  & 0.1 & \hspace{0.2cm}  23 \hspace{0.2cm} & 2.4  & 1\\
A2 & $512^3$ & $\infty$  & 0.1 & \hspace{0.2cm}  50 \hspace{0.2cm} & 2.4  & 1\\
B & $256^3$ & $\infty$ & 0 &  90 & 0.2 & 75\\
D & $256^3$ & $\infty$ & 0.8 &  30.7 & 1.9  &1\\
\hline
\hline
\end{tabular}

\caption{$\delta B/|\langle {\bf B} \rangle|$ is the ratio of the rms
fluctuating field to the mean field, $H_m$ is the magnetic helicity
divided by the maximum possible magnetic helicity at that level of
magnetic energy, $l_B/l_d$ is the ratio of outer scale to inner scale,
and $\alpha$ is the coefficient in equation~(\ref{eq:tr}).
\label{tab:t1} }
\end{table}

We take a snapshot of the magnetic field in each simulation
and introduce 2,000 pairs of field-line tracers whose initial
separation~${\bf r}_0$ is perpendicular to the local field.  We use
linear interpolation to obtain the magnetic field between grid points
and employ second-order Runge-Kutta to integrate field-lines.
To improve the statistics in simulation~A1,
we use 2,000 field lines in each of
seventeen snapshots of the field separated in time
by an interval~$0.4l_B/u$, where~$u$ is the rms velocity.
In simulations~A2 and~A2$_{\rm rp}$,
we use 20,000 field lines in each of five snapshots
of the magnetic field separated in time by an interval~$0.2l_B/u$.
We iteratively reduce the length step in the
field-line integrations to achieve convergence.

\begin{figure}[h]
\vspace{8.5cm}
\includegraphics{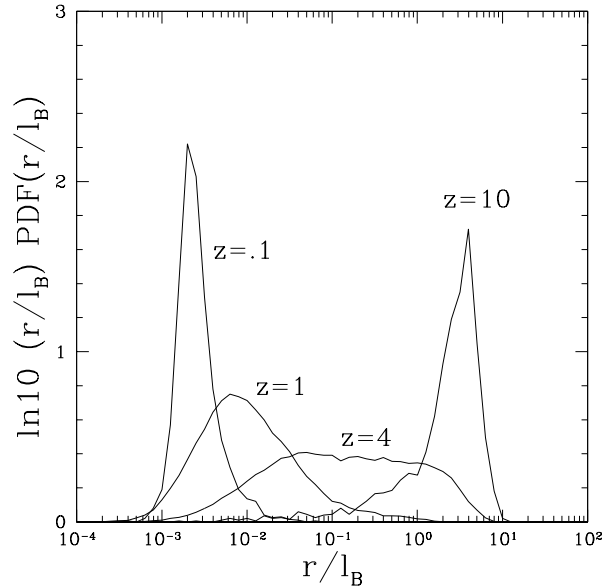}
\caption{The probability distribution function of field-line separations
in simulation~A1 with $r_0 = 2^{-9} l_B$.
\label{fig:f1}}
\end{figure}

The probability distribution function (PDF) of field-line separations
in simulation~A1 for $r_0 = 2^{-9}l_B$ is plotted in
figure~\ref{fig:f1} for different values of the distance~$z$ a
field-line pair is followed along the magnetic field.
[The probability is per unit $\log_{10}(r/l_B)$, so that the
probability that~$r$ lies in some interval is proportional
to the area under the plotted curves.]  The PDF is
highly non-Gaussian, and the tail of the distribution dominates the
growth of $\langle r \rangle$ when $l_d < \langle r \rangle < l_B$.
The maximum kurtosis $\langle r^4 \rangle/ \langle r^2 \rangle^2$ is
$\sim 500$ and occurs for $z\simeq 0.2$.  We find that decreasing
$r_0$ increases the maximum kurtosis.

\begin{figure}[h]
\vspace{8.5cm}
\includegraphics{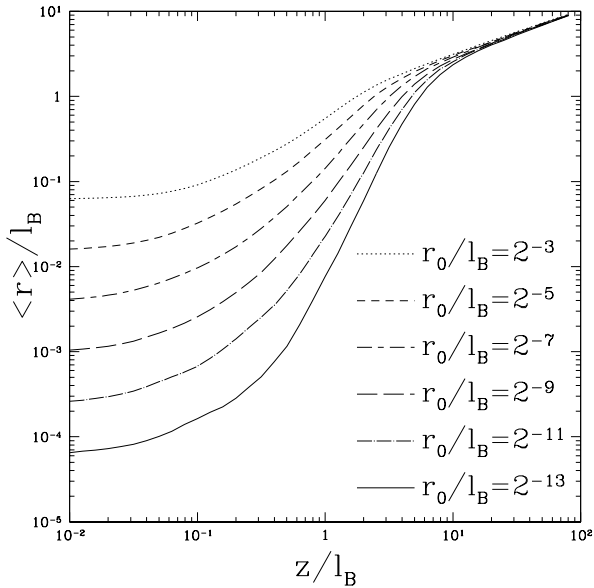}
\caption{Mean field-line separation $\langle \triangle r\rangle$ as a function
of distance $l$ travelled along the magnetic field in simulation~A
for different values of~$r_0$.
\label{fig:f2}}
\end{figure}

The growth of $\langle r \rangle $ with $z$ in simulation~A1 is shown
in figure~\ref{fig:f2} for several values of~$r_0$. There are three
stages of growth similar to those described in previous theoretical
treatments of field-line divergence~\cite{nar01,ski74}: (1) an initial
stage of exponential growth when $r_0 \ll l_d$, (2) approximate
power-law growth with $r\propto z^a$ for $\langle r \rangle \ll l_B$,
and (3) $\langle r \rangle \propto z^{1/2}$, for $\langle r \rangle >
l_B$.  However, some aspects of our results differ from previous
studies.  Within a single simulation, $a$~increases with
decreasing~$r_0$, which is probably related to the increasing
prominence of the tail of the PDF as $r_0$ is decreased.  In addition,
stage~2 with $\langle r \rangle \propto z^a$ begins for $\langle r
\rangle < l_d$, perhaps because the field-line pairs with largest~$r$,
which dominate the growth of~$\langle r \rangle$, satisfy $r> l_d$
before $\langle r \rangle > l_d$.

\begin{figure}[h]
\vspace{9cm}
\includegraphics{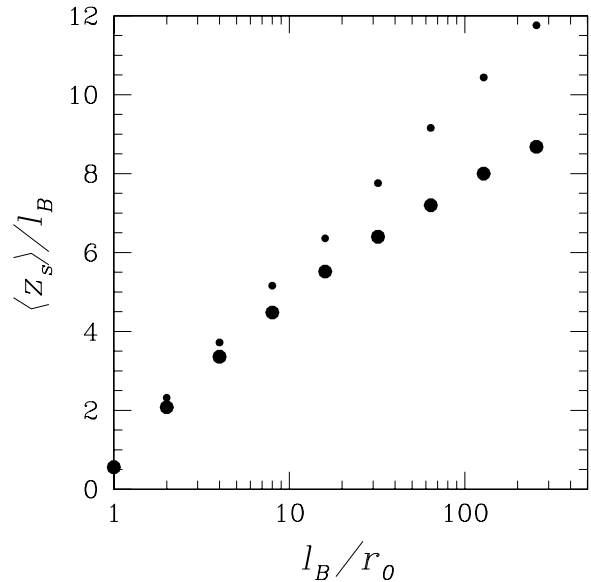}
\caption{The average distance $\langle z_s \rangle$
that a field-line pair must be followed before separating by a
distance~$l_B$ as a function of
initial field-line separation~$r_0$ for simulations~A1 (small
circles) and~A2 (large circles).
\label{fig:f3}}
\end{figure}

\begin{figure}[h]
\vspace{8.5cm}
\includegraphics{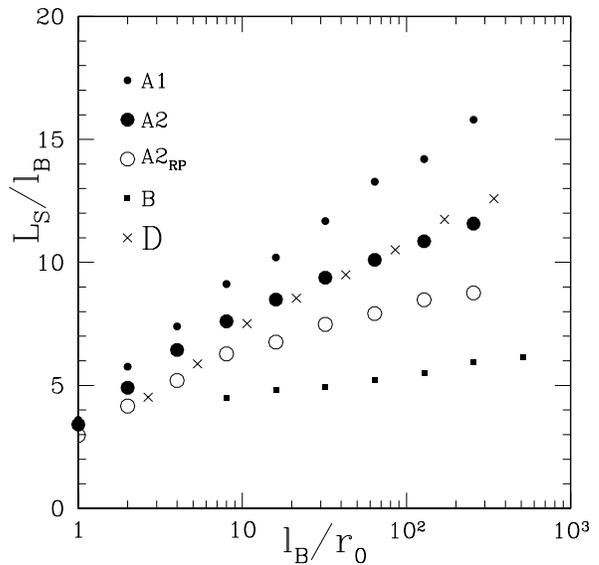}
\caption{The dependence of $L_{\rm S}$ on $r_0$.
\label{fig:f4}}
\end{figure}

We seek to test the qualitative prediction of~\cite{jok73,ski74,nar01}
that~$\langle z_s \rangle$
and~$L_{\rm S}$ asymptote to a value of order a few~$l_B$  as~$r_0$ is
decreased towards~$l_d$ in the large-$l_B/l_d$ limit. In figure~\ref{fig:f3}, we
plot~$\langle z_s \rangle$ for simulation~A1 and
simulation~A2. The lower-resolution data of
simulation~A1 suggest the scaling~$\langle z_s \rangle \propto \ln
(l_B/r_0)$ for $l_d < r_0 <0.25 l_B$, in contradiction to the
theoretical treatments. On the other hand, for simulation~A2,
the curve through the data is concave downward for~$l_d < r_0 <
l_B$. Moreover, figure~\ref{fig:f3} shows that the
simulation~A1 data, and probably also the A2~data, have not converged
to the high-Reynolds-number values of~$\langle z_s \rangle$ 
for~$l_B/16 < r_0 < l_B$, values of~$r_0$ that are within the
inertial ranges ($l_d$ to $l_B$) of both simulations. Also,
the slope $d\langle z_s \rangle/d(\ln(l_B/r_0))$ for both~$r_0<l_d$
and~$r_0<l_B/10$ 
decreases significantly when $l_B/l_d$ is doubled. A comparison
of the data for~A1 and~A2 thus suggests that in the large-$l_B/l_d$ limit
$\langle z_s \rangle$ asymptotes to a value of order several~$l_B$
as~$r_0$ is decreased towards~$l_d$, as in~\cite{jok73,ski74,nar01}.
The same comments apply to
the data for~$L_{\rm S}$, which are plotted in figure~\ref{fig:f4}.

We note that for $r_0 = l_B$, $\langle z_s \rangle$ and~$L_{\rm S}$ are
by definition~0.  The numerical-simulation data points that appear to
be plotted above~$l_B/r_0 = 1$ actually correspond to~$r_0$ just slightly
smaller than~$l_B$, indicating that $\langle z_s \rangle$
and~$L_{\rm S}$ are discontinuous at $r_0 = l_B$ in the numerical
simulations. The reason is that for $r_0 $ just slightly less than
$l_B$, some fraction of the field line pairs are initially converging
and must be followed a significant distance before they start to
diverge.

The thermal conductivity $\kappa_T$ in galaxy-cluster plasmas scales
approximately like the diffusion coefficient of thermal
electrons~\cite{rec78,kro83}.  For clusters, $l_B/l_d \simeq
l_B/\rho_i \simeq 10^{13}$, where~$\rho_i$ is the proton
gyroradius~\cite{qua98}.  In the large-$l_B/l_d$ limit, the numerical
simulations and theoretical models indicate that $L_{\rm S}$
asymptotes to a value of order several~$l_B$ as~$r_0$ is decreased
towards~$l_d$, and $L_{\rm S}$ is not expected to increase appreciably
as $r_0$ is further decreased from~$l_d=\rho_i$ to~$\rho_e$. Thus,
$L_{\rm S}(\rho_e) \simeq L_{\rm S}(l_d)$.  We take simulations~A1
and~A2 to be our best models of a galaxy-cluster magnetic field. We
find that $L_{\rm S}(l_d) \simeq 11 l_B$ in~A1 with~$l_B/l_d=23$, and
$L_{\rm S} (l_d) \simeq 10-11 l_B$ in~A2 with~$l_B/l_d = 50$. If~$l_B$
is redefined so that~$2\pi/l_B$ (instead of~$\pi/l_B$) equals the
maximum of~$kE_b(k)$, then $L_{\rm S}(l_d) \simeq 7l_B$ in~A1
and~$L_{\rm S}(l_d)\simeq 6.5l_B$ in~A2\cite{cha03}. Since the
definition of~$l_B$ contains an arbitrary constant of order unity, it
is not clear which of the two definitions leads to the more accurate
prediction of~$\kappa_T$.  Given this uncertainty, extrapolating these
results to the large-$l_B/l_d$ limit suggests that $L_{\rm S} (l_d)
\sim 5-10 l_B$ in intracluster plasmas. Diffusion along the magnetic
field can be suppressed by magnetic mirrors~\cite{cha98,mal01} and
wave pitch-angle scattering~\cite{pis98}, but in clusters the Coulomb
mean free path is sufficiently short that neither of these effects is
very important.~\cite{nar01,cha03} Thus, the parallel diffusion
coefficient of thermal electrons is comparable to the thermal-electron
diffusion coefficient in a non-magnetized plasma,~$D_0$.  In the
static-magnetic-field approximation, equation~(\ref{eq:d2}) thus
implies that $D/D_0 \sim 0.1-0.2$ for thermal electrons in
intracluster plasmas, and that $\kappa_T$ is reduced by a factor
of~$\sim 5-10$ relative to the Spitzer thermal conductivity~$\kappa_{\rm S}$ of a
non-magnetized plasma. 
Heating of intracluster plasma from thermal conduction 
with $\kappa_T\sim 0.1-0.2\kappa_{\rm S}$ would be
sufficient to balance radiative cooling in a some but not
all clusters~\cite{fab02,zak03}.
More work is needed to determine the validity
of the static-field approximation in clusters, to determine the
factors of order unity that have been neglected in our
phenomenological treatment of thermal conduction, and to clarify the
effects of turbulent diffusion and turbulent resistivity on heat
conduction in strong MHD turbulence.

We thank Steve Cowley and Alex Schekochihin for valuable discussions.
This work was supported by NSF
grant AST-0098086 and DOE grants DE-FG02-01ER54658 and
DE-FC02-01ER54651 at the University of Iowa, with computing resources
provided by the National Partnership for Advanced Computational
Infrastructure and the National Center for Supercomputing
Applications.

\end{document}